\documentclass[conference]{IEEEtran}
\IEEEoverridecommandlockouts
\usepackage{cite}
\usepackage{amsmath,amssymb,amsfonts}
\usepackage{algorithmic}
\usepackage{algorithm}
\usepackage{graphicx}
\usepackage{textcomp}
\usepackage{xcolor}
\usepackage{booktabs}
\usepackage{subfigure}
\usepackage{flushend}

\usepackage[colorlinks=true, urlcolor=blue]{hyperref}
\usepackage{breakurl}

\def\BibTeX{{\rm B\kern-.05em{\sc i\kern-.025em b}\kern-.08em
    T\kern-.1667em\lower.7ex\hbox{E}\kern-.125emX}}
\begin{document}
\newcommand{\name}{Pier}
\title{\name{}: Efficient Large Language Model pretraining with Relaxed Global Communication\\
}

\author{\IEEEauthorblockN{Shuyuan Fan}
\IEEEauthorblockA{\textit{Department of Electrical and Computer Engineering} \\
\textit{Rutgers University}\\
New Brunswick, United States\\
sf850@scarletmail.rutgers.edu}
\and
\IEEEauthorblockN{Zhao zhang}
\IEEEauthorblockA{\textit{Department of Electrical and Computer Engineering} \\
\textit{Rutgers University}\\
New Brunswick, United States\\
zhao.zhang@rutgers.edu}
}

\maketitle

\begin{abstract}
Global communication, such as all-reduce and all-gather, is the prominent performance bottleneck in large language model (LLM) pretraining.
To address this issue, we present \name{}, an efficient and scalable optimizer with relaxed global communication.
\name{} is built upon DiLoCo, which leverages an inner optimizer within groups of processors and an outer optimizer that requires global communication.
To preserve the convergence and model performance, \name{} incorporates two key techniques for the outer optimizer: momentum warmup and momentum decay.
\name{} employs an efficient and scalable system architecture to enable complex parallelization strategies in LLM pretraining.
We examine the model performance and runtime reduction of \name{} using the GPT model family (e.g., small, medium, XL, and 7B) and the OpenWebText dataset with a suite of thirteen downstream tasks.
With data parallel strategy, \name{} speeds up GPT-2 XL training by up to 2.7x-3.7x on 256 NVIDIA A100 GPUs and 1.2x-1.9x on 64 GH200 Superchips, respectively, without degradation of validation loss or downstream task performance.
With data parallel and tensor parallel, \name{} reduces the time cost GPT-2 7B model training by 54.5\% on 128 A100s.
\end{abstract}

\section{Introduction}
Major stakeholders are designing large language models (LLMs) with ever-increasing size, driven by the scaling law~\cite{kaplan2020scaling}, which states that the model's capability improves with scale.
The state-of-the-art LLMs~\cite{BrownMann2020-gpt3, jiang2024megascale, grattafiori2024llama} are trained with tens of thousands of high-end graphic processing units (GPUs) in complex distributed strategies, including data parallel~\cite{li2020pytorch}, tensor parallel~\cite{shoeybi2019megatron}, pipeline parallel~\cite{huang2019gpipe, narayanan2019pipedream}, and context parallel~\cite{liu2023ring}.
Communication is the prominent bottleneck in LLM pretraining, as simply evidenced by the low scaling efficiency of 42.7\% and 34.6\% for GPT-2 1.5B training with 32 NVIDIA A100s and 64 NVIDIA GH200 superchips using the AdamW optimizer with standard data parallel, respectively (see AdamW curves in Figure~\ref{fig:scale-dual}).

\begin{figure}[t]
  \centering
  \includegraphics[width=1\columnwidth]{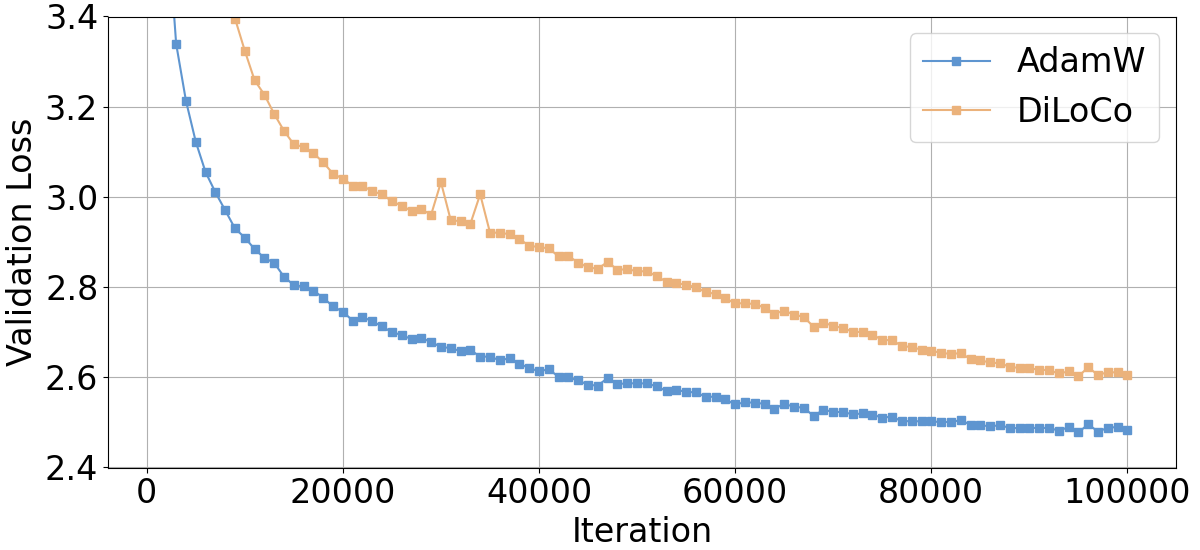}
  %\vspace{0.1in}
  \caption{\fontsize{9pt}{11pt}\selectfont
  Validation Loss Comparison between AdamW (8 GPUs, Fully Synchronized) and DiLoCo (8 Groups, 1 GPU per Group) During the Pretraining of GPT-2 XL Model.}
  \label{fig:problem_setup}
  %\vspace{0.1in}
\end{figure}
To reduce communication costs and boost scaling efficiency for LLM pretraining, researchers have proposed algorithms that fall into two categories.
The first category lowers the communication volume via gradient sparsity~\cite{lin2018deep, zheng2025radius} and quantization~\cite{faghri2020adaptive}.
%The second category overlaps communication with back-propagation~\cite{li2020pytorch, wang2024domino}.
And the second category reduces the global communication frequency by exploiting optimizers such as local SGD~\cite{stich2018local}.
With the hard constraints of model performance (i.e., validation loss and downstream task performance), methods in the first category exhibit modest improvements of 10-30\%~\cite {song2023optimusccefficientlargenlp, zheng2025radius}.
Although the methods with relaxed global communication show great potential in pretraining acceleration, they suffer from performance degradation.
Among these methods, DiLoCo~\cite{douillard2023diloco} introduces a two-layer optimization framework that partitions GPUs into groups.
It utilizes an inner optimizer that only requires intra-group communication and an outer optimizer that requires inter-group (i.e., global) communication.
However, our empirical evaluation in Figure~\ref{fig:problem_setup} shows degraded validation loss and downstream task performance using the GPT-2 XL model and OpenWebText dataset~\cite{Gokaslan2019OpenWebText}.

Our analysis of DiLoCo reveals that the gradient oscillation at the switching point from AdamW to DiLoCo is the root cause of performance degradation. 
This insight motivates the design of \name{}, which incorporates two key techniques: momentum warmup and momentum decay for the outer optimizer.
The momentum warmup technique tracks the model changes as the outer optimizer's gradients, for every few iterations during the so-called ``lazy start'' phase in DiLoCo, where the model is updated by using the same AdamW as baseline. 
After the ``lazy start'' phase, the momentum decay technique exploits a fixed schedule to gracefully migrate from AdamW to DiLoCo.
To reduce the memory overhead in the outer optimizer, \name{} offloads the model and optimizer states to host memory during inner loops. 
\name{} enables the overlay of data parallel over tensor parallel to support larger models, where the outer step parallelizes communication and model updates.

%While the original DiLoCo paper introduced a "lazy start" technique--an initial phase of standard data parallel pretraining before switching to DiLoCo, it significantly underestimated its role, claiming a trivial impact on convergence. Our key insight is that the ineffectiveness of this "lazy start" is from the failure to address the instability that occurs at the switching point, which caused a waste of training iterations. We believe the cause of this instability is the sudden mismatch on the outer optimizer state (the momentum). This insight directly enables our key idea: warm up the state of the outer optimizer in the initial phase to reduce the mismatch at the switching point. Hereby, we proposed Pier, a technique for fault tolerance system with lazy start mechanism. Our primary contributions in this work are as follows:

\name{} is implemented upon the Megatron-LM framework. 
We evaluate the design effectiveness of \name{} using the GPT-2 model family, including small (125M), medium (345M), XL (1.5B), and a 7B model with the OpenWebText dataset.
We examine the convergence and downstream task performance of \name{} using a suite of thirteen tasks.
With the model performance guarantee, we conduct five studies on the scaling patterns of \name{}.
We first perform a weak scaling study to identify the global batch size boundary of \name{} with a fixed token budget and group sizes.
The second study explores the convergence boundary when increasing the synchronization interval of the outer optimizer. 
We then perform a strong scaling study to determine the scaling efficiency boundary of \name{} with fixed global batch sizes and the number of groups. 
In the fourth experiment, we vary the group size (i.e., the number of GPUs in a group) and fix the global batch size to study the impact of the number of groups on training speed. 
In the fifth study, we perform a mixed parallelism scaling experiment to examine the scalability of \name{} for larger models. 
%\zhao{Shuyuan, add more experiments that I missed.}

We run the experiments on two GPU clusters: NERSC Perlmutter and TACC Vista.
\name{} shows comparable validation loss and downstream task performance with the baseline AdamW optimizer in standard data parallelism for GPT-2 small, medium, and XL.
On Perlmutter, \name{} speeds up GPT-2 small, medium, and XL training by 1.7x, 2.6x, and 2.7x with the maximum data parallel scale that guarantees model performance. 
If global communication is further relaxed, the speedup for GPT-2 XL can be as high as 3.7x on 256 A100s.
On Vista, \name{} is effective on GPT-2 XL with data parallel. 
The speedup is 1.2-1.9x on 128 GH200s.
For GPT-2 7B training with data parallel and tensor parallel, \name{} exhibits a speedup of 2.2x on 128 A100s.
%\zhao{Shuyuan, add more experiment results.}

%Our implementation is built upon the Megatron-LM framework. We evaluate our approach on GPT-2 models, from GPT-Small (125M parameters) to GPT-XL (1.5B parameters). Our experiments were conducted on two distinct clusters: the vista cluster with single NVIDIA H100 GPU per node where nodes connected by the newest infiniband, and the Perlmutter cluster with 4 NVIDIA A100 GPUs per node where nodes connected by slingshot. Our speedup benchmarks are conducted with different scope: with fixed number of subgroups and varying number of subgroups, where each subgroup is a group of GPUs that communicate per iteration. The code will be available upon acceptance.
The contributions of this paper are:
\begin{itemize}
\item The momentum warmup and momentum decay techniques that allow \name{} to achieve comparable model performance as AdamW. 
\item The efficient \name{} design that combines data parallel and tensor parallel.
%\item The 2.7-3.7x and 1.2-1.9x speedup for GPT-2 XL model pretraining on 256 A100s and 128 GH200s, repsectively.
\item The comprehensive model performance evaluation with thirteen downstream tasks.
\item The quantified end-to-end LLM pretraining time improvement over two production supercomputers with different GPUs and interconnect architectures.
\end{itemize}

%\outline{
%Experimental methodology and artifact availability. 
%Clearly specify the key experimental / simulation infrastructure and methodological details. 
%Support the experimental methodology choices (e.g., cite that the most relevant and most recent prior works have evaluated their ideas using similar methodology). 
%}

%\outline{Optional. Limitations of the proposed approach. Almost all scientific contributions have limitations and scope for improvement. Clearly articulate all the major limitations of the proposed approach and identify conclusions that are sensitive to specific assumptions made in the paper.}

\section{Background}
In this section, we introduce the basic concepts of distributed LLM pretraining strategies, the communication bottleneck due to bandwidth hierarchy, and the structure of the DiLoCo optimizer. 
\subsection{Distributed Model Training}
Modern LLM pretraining exploits complex parallelism strategies. 
%including expert-parallel, data-parallel, pipeline-parallel, context-parallel, and tensor-parallel. 
3D parallelism partitions a model across processors by layering data parallelism (DP) over pipeline and tensor parallelism (PP and TP).
GPUs are partitioned into groups (Data Parallel Rank), and each group hosts a complete copy of the model.
Inside each group, the model is spread across nodes by layers to form the pipeline stages;
Inside each node, layers (tensors) are evenly distributed across multiple GPUs, a technique referred to as tensor parallelism.
With the data parallel approach, the only communications are the broadcast at the start of training and the gradient exchange in each iteration.
In contrast, 3D parallelism requires intra- and inter-node communication for both forward and backward computation.

With the need for larger context in LLMs, 
long sequence support is introduced as a fourth dimension. 
Researchers have designed context parallelism (CP) with various partitioning and communication strategies over the query, key, and value matrices~\cite{li2022sequence, liu2024blockwise, liu2023ring}.
In addition, mixture-of-expert (MoE) models can leverage the expert parallelism (EP), where each expert is a standalone LLM.
%\autoref{fig:ringattention} exemplifies ring attention, which employs an outer loop of query blocks (across GPUs) and inner loops for key-value blocks (within a GPU).
%Among existing methods, ring attention achieves the lowest memory overhead by using an outer loop of query blocks (across GPUs) and inner loops for key-value blocks (within a GPU), as shown in \autoref{fig:ringattention}.
%However, ring attention suffers from the low bandwidth in inter-node communication, i.e., with Infiniband across four nodes (12.5~GB/s one-way bandwidth among 32 A100s), it requires a 24.5$\times$ larger block size to maximize the GPU utilization compared to a single DGXA100 server (300~GB/s one-way bandwidth among 8 A100s).
%This limited scalability is attributed to the low bandwidth between nodes, where the communication overhead cannot be amortized by the computation. 
These distributed LLM pretraining strategies complicate the design of optimizers, as the models and optimizer states are spread across GPUs.
In \name{}, we design an efficient parallel strategy for the communication in the outer optimizer under TP and DP.

\subsection{Bandwidth Hierarchy in Networks}
State-of-the-art supercomputers are equipped with networks that feature hierarchical bandwidth. 
Within a compute node, GPUs are usually connected via NVLink, Infinity Fabric, or Xe Link. 
E.g., NVLink 4.0 and 5.0 support a bandwidth of 900 GB/s and 1.8 TB/s, respectively.
These supercomputers deploy Infiniband or Slingshot to connect a massive number of compute nodes. 
E.g., the Infiniband NDR offers 100~GB/s bandwidth and HDR offers 50~GB/s.
The global communication over the interconnects is the scaling bottleneck for LLM pretraining at extreme scale. 
This is because the all-reduce or all-gather is needed in every iteration.
However, there exists an opportunity for lower communication cost if an optimizer can efficiently leverage the high bandwidth in the intra-node links.
This hierarchical bandwidth in the supercomputer networks motivates the design of \name{}.

%Modern supercomputers are equipped with low-latency and high-bandwidth interconnects in topologies such as FatTree and DragonFly.
%These heterogeneous topologies have hierarchical latency and bandwidth.
%In FatTree, communication between two nodes under the same edge switch takes just one hop;
%in contrast, two nodes under different core switches may take five hops, although the core switch has a higher bandwidth, e.g., 200~Gbps rather than 100~Gbps.

%Although software-level optimizations are crucial, the underlying hardware also plays a significant role in the communication cost. For example, in modern supercomputers used for training large language models, machines are composed by multiple compute nodes, these nodes are connected in different topological patterns. The nodes are connected by Infiniband, which has a good latency and bandwidth. Inside each node, the gpus are connected by NVLink, which is faster than Infiniband. When people use multiple nodes for a large scale problem, the communication cost significantly increases than the single node run. \par
%The speed of internode connections varies. For example, the perlmutter by nersc uses slingshot link, which has some performance issues. Similarly, some computers like vista uses newest infiniband, which is faster than old inifiniband on polaris.

\subsection{DiLoCo}
DiLoCo was motivated by the federated learning scenario, where the network bandwidth is extremely limited across different geolocations. 
It forms groups based on the geographical locations of the workers, exploits a two-layer optimization approach.
%It first partitions the data parallel (DP) ranks into several groups, and splits the global batch size across the DP ranks, following the conventional data parallel training approach. 
%The training process consists of a two-level loop. 
In the inner loop, the model is trained with AdamW leveraging standard DP. 
The gradient synchronization is within each group, thus eliminating the global communication, which results in model parameter divergence across groups.
Thus, for every few iterations (i.e., inner loop interval), the outer optimizer needs to gather the model weights from all groups and treat the model changes as gradients.
Then the outer optimizer takes an optimization step to update the model and synchronize the model across groups.
Empirically, Nesterov SGD shows better performance than SGD, momentum SGD, and AdamW as an outer optimizer~\cite{douillard2023diloco}.

%After each selected iteration interval, in the outer loop, weight changes of the models across the groups are collected as the gradient of the selected outer optimizer. 
%This outer optimizer will synchronize the model by taking an optimization step, where gradients are applied to the model weights from the last synchronization. 
%Their empirical results show the Nesterov optimizer outperformed other optimizers, including SGD, SGD-M, and AdamW. 
%Note that the outer optimizer maintains a momentum term when the selected optimizer is SGD-M, Nesterov or AdamW.
\section{Related Works}
Researchers have studied various methods to reduce the communication overhead in LLM pretraining.
These methods can be classified into two categories: lowering communication volume and reducing the frequency of communication.

\subsection{Volume Reduction}
The data compression approach is one of the most effective approaches for reducing communication cost in distributed deep learning. 
The early work~\cite{seide2014} introduces 1-bit SGD with an error feedback mechanism to compensate for errors. 
It maintains convergence while reducing communication cost. 
Another technique used in gradient compression is sparsity, which communicates gradients with large magnitudes rather than transmitting all the gradients. 
EF21~\cite{richtarik2021ef21} improves the classical error feedback mechanism, achieving both theoretical and practical convergence. 
Variance-based Gradient Compression~\cite{tsuzuku2018variance} only transmits gradient entries that are unambiguous, while keeping ambiguous components local, effectively balancing compression and accuracy. 
Deep Gradient Compression (DGC)~\cite{lin2018deep} combines several practical mechanisms, including warm-up, local gradient clipping, momentum correction, and sparsification into a unified framework. 
Moving beyond quantization and sparsification, PowerSGD~\cite{vogels2019powersgd} compresses gradients through low-rank matrix approximation, providing a communication-efficient approach suitable for large-scale machine learning systems. 
%More recently, EF21~\cite{richtarik2021ef21} improves the classical error feedback mechanism, achieving both theoretical and practical convergence. 
%\zhao{summarize their limitation in LLM pretraining.}
%Many compression methods show stable convergence under mild compression. However, pushing compression ratio high often risks convergence or accuracy.
In practice, with the constraint of model performance (i.e., validation loss and downstream task performance), compression techniques~\cite{zheng2025radius, song2023optimusccefficientlargenlp} can reduce the time cost by 10-20\%.
%However, in practice, aggressive compression leads to faster training but unstable convergence, while mild compression remains stable but only brings a modest speedup. 
%This trade-off is the key limitation of such approaches.

\subsection{Reducing Communication Frequency}
Another series of works aims at reducing the frequency of communication. 
Local SGD~\cite{stich2018local} updates models on local workers every a few iterations, and synchronizes the diverged models across different workers via averaging model weights. 
This idea is widely adopted in federated learning. 
The later work~\cite{gorbunov2021local} implements a unified theoretical framework for Local SGD and other federated learning approaches, and provides a convergence analysis under both \textit{i.i.d.} and \textit{non-i.i.d.} data settings.

DiLoCo~\cite{douillard2023diloco} shows great potential in global communication reduction for pretraining of the decoder-only LLMs by partitioning GPUs into groups and synchronizing periodically. 
A follow-up research ~\cite{charles2025communication} presents a scalability analysis of DiLoCo, claiming better model performance with DiLoCo than AdamW for larger models.
However, this study only shows the evidence of lower validation loss, which overlooks the downstream tasks. 
SparseLoCo~\cite{sarfi2025communication} enhances communication efficiency further by applying top-k sparsification and quantization during the global synchronization step, achieving sparsity as low as \~1-3\%. 
It outperforms the AdamW and DiLoCo baseline in communication volume, validation loss, and three downstream tasks.
Streaming-DiLoCo~\cite{douillard2025streaming} improves DiLoCo by (1) synchronizing only subsets of the model parameters to reduce the peak communication demand; (2) overlapping communication and computation; (3) applying low-precision quantization for outer optimizer updates. 
While these approaches can reduce communication cost, the convergence and model performance is only evaluated with a few metrics. 
These studies also lack quantified measurements of communication time reduction in end-to-end pretraining on production supercomputers.

In contrast, \name{} places convergence and model performance as the top priority. 
\name{} integrates two techniques to preserve training convergence and downstream task performance, which is the hard constraint for the study of scalability and training time reduction.

\section{Algorithm and System Design}
\label{sec:design}
In this section, we introduce the two key techniques that preserve the model performance of \name{}. 
We also present the scalable design of \name{} in LLM pretraining, incorporating both data parallel and tensor parallel approaches.

\begin{algorithm}[t]
   \caption{MomentumWarmup}
   \label{alg:momentum_warmup}
    \textbf{Input:} Training Iteration $T$ \\
    \textbf{Input:} Warmup percentage $p$ \\
    \textbf{Input:} Synchronization interval $r$ \\
    \textbf{Initialize:} Momentum coefficient $\mu$ \\
    \textbf{Initialize:} Model parameters $\theta$ \\
    \textbf{Initialize:} Momentum buffer $M$ \\
     \textbf{Output:} Updated model parameters $\theta_{pT}$ \\
     \textbf{Output:} Momentum buffer $M$
\begin{algorithmic}[1]
   \FOR{$t = 1$ {\bfseries to} $pT$}
        \STATE Compute local gradient $\nabla_{\theta}$
        \STATE $\theta_t \leftarrow OptimizerUpdate(\theta_{t-1}, \nabla_{\theta})$
             \IF{$t \bmod r = 0$}
                \STATE $\Delta\theta \leftarrow \theta_t - \theta_{t-r}$
                \STATE $M \leftarrow \mu M +  \Delta\theta$
             \ENDIF
    \ENDFOR
\end{algorithmic}
\end{algorithm}

\subsection{Momentum Warmup}
In the original DiLoCo study, exploiting the so-called ``lazy-start'' technique, where training begins with AdamW then switches to DiLoCo, leads to lower model performance than the AdamW baseline~\cite{douillard2023diloco}. 
The cause of the performance degradation is the training instability during the transition from AdamW to DiLoCo.
To mitigate the instability issue, we design the momentum warmup technique.
During the ``lazy-start'' phase, e.g., the first 10\% of the pretraining, \name{} uses AdamW in standard data parallelism.
This is to leverage AdamW's capability of quickly escaping local minima in the early stage.
At the same time, \name{} accumulates the momentum by differencing the models' states with the same interval as the outer optimizer.
Please note that, in the ``lazy-start'' phase, \name{} only accumulates the momentum without applying it for model updates.

%During the early 10\% of the pretraining, we useuseandard data parallelism to take advantage of its fast early convergence. Although the DiLoCo-like outer optimizer is not yet initialized at this stage, we still want prepare it for later use. When the outer optimizer is Nesterov-SGD, SGD-M, or AdamW, it uses an accumulated momentum in the update formula, which suffers from a severe unmatch if it's randomly initialized after the lazy start stage.
Algorithm~\ref{alg:momentum_warmup} illustrates the momentum warmup technique. 
The default value of the momentum coefficient is $\mu=0.9$.
With momentum warmup, Pier initializes the outer optimizer with the accumulated momentum.

%This motivates our design of Pier. In the early 'lazy start' stage, Pier records the change of model parameters at fixed intervals, which should be the same interval that will be used later. We regard this change of model parameters as the gradient of outer optimizer and update its momentum with selected coefficient $\mu=0.9$. When training switches to into DiLoCo, Pier initializes the outer optimizer with the accumulated momentum prepared during the lazy start stage.

\subsection{Momentum Decay}
\label{sec:design:decay}
After switching to DiLoCo, \name{} incorporates the momentum decay technique for graceful transition.
The intuition is to leverage the faster empirical convergence rate of AdamW compared to SGD variants with momentum, as AdamW provides a normalized and adaptive estimate of the gradient direction. 
Thus, \name{} exploits a momentum schedule after the optimizer transition. 
The details are shown in Algorithm~\ref{alg:decay}.

Specifically, \name{} runs the ``lazy start'' phase for the first 10\% of the overall training. 
Then \name{} switches to DiLoCo.
From 10\% to 15\% of total training steps, the momentum coefficient $\mu$ is set to 0.99 to avoid sharp changes in velocity during the unstable stage.
From 15\% to 20\%, $\mu$ decays to 0.95.
After 20\% of the total steps, $\mu$ remains at 0.9, matching the recommended hyperparameter setting for the DiLoCo outer optimizer.
In practice, we find combining the momentum warmup and decay techniques is effective in GPT model pretraining, including GPT small, medium, and XL (see \S\ref{sec:expr:convergence}). 

%After the lazy start phase, we introduce a momentum decay mechanism to smoothly initialize the DiLoCo-like outer optimizer in Pier. Instead of using a fixed value momentum coefficient value, we design a scheduled decay for the momentum coefficient $\mu$, allowing the optimizer to adapt gradually. To be specific, during the first 10\% of training (the lazy start phase), Pier only accumulates outer momentum but never uses it. From 10\% to 15\% of total training steps, $\mu$ is set to 0.99 to avoid rapid changes in momentum during this relatively unstable stage; from 15\% to 20\%, $\mu$ decays to 0.95. After 20\% of the total steps, $\mu$ remains at 0.9, matching the recommended hyperparameter setting for the DiLoCo outer optimizer.

\begin{algorithm}[t]
   \caption{Pier}
   \label{alg:decay}
    \textbf{Input:} Train Iterations $T$ \\
    \textbf{Input:} Warmup Percentage $p$ \\
    \textbf{Input:} Synchronization interval $r$ \\
    \textbf{Initialize:} Momentum coefficient $\mu$ \\
    \textbf{Initialize:} Model parameters $\theta$ \\
    \textbf{Initialize:} Momentum buffer $M$ 
\begin{algorithmic}[1]
\STATE $\theta , M \leftarrow$ MomentumWarmup($T,p,r$)
\STATE Initialize data parallel sub-groups: $\{\mathcal{G}_1, \mathcal{G}_2, \dots, \mathcal{G}_k\}$ 
\FOR{$t = pT$ {\bfseries to} $T$}  
    \IF{$t \bmod r \ne 0$} 
        \FOR{each subgroup $\mathcal{G}_i$}
            \STATE Compute local gradient $\nabla_{\theta}$
            \STATE $\theta_t \leftarrow OptimizerUpdate(\theta_{t-1}, \nabla_{\theta})$
        \ENDFOR
    \ELSE
        \STATE $\Delta\theta \leftarrow \theta_t - \theta_{t-r}$  
        \STATE $\Delta\theta \leftarrow AllReduce(\Delta\theta)$  \COMMENT{across all ranks}

        \IF{$0.1T \le t < 0.15T$}
        \STATE $\mu \leftarrow 0.99$
        \ELSIF{$0.15T \le t < 0.2T$}
        \STATE $\mu \leftarrow 0.95$
        \ELSE
        \STATE $\mu \leftarrow 0.9$
        \ENDIF
        \STATE $lr \leftarrow outer\_lr\_scheduler(t)$ 
        \STATE $M \leftarrow \mu M + \Delta\theta$  
        \STATE $\theta_t \leftarrow \theta_{t-r} + lr *(M *\mu + \Delta\theta)$ 
    \ENDIF
\ENDFOR
\end{algorithmic}
\end{algorithm}

\begin{figure}[h]
  \centering
  \includegraphics[width=1\columnwidth]{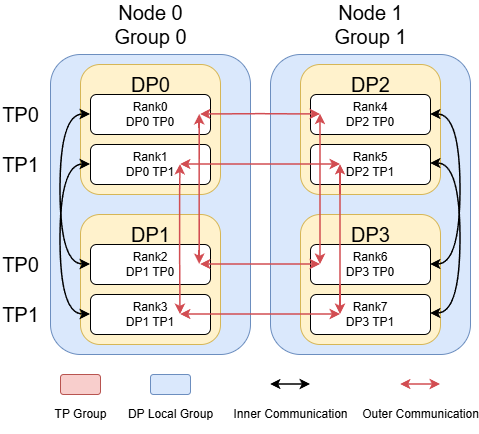}
  %\vspace{0.1in}
  \caption{Illustration of Inner- and Outer-Communication in a 2D Parallel Training Setup with Data Parallel Size of 4 and a Tensor Parallel Size of 2. There are two local communication groups, each located on a separate compute node.}
  \label{fig:tp_figure}
  %\vspace{0.1in}
\end{figure}
%\shuyuan{outer lr scheduler is needed in algorithm, but content is in implementation}

\subsection{2D Parallelism}
The design of vanilla DiLoCo ~\cite{douillard2023diloco} replaces full data parallel gradient synchronization with a low-communication alternative. 
However, in large-scale training, data parallelism is often combined with tensor parallelism.
The complex parallelism in LLM pretraining complicates the efficient and scalable \name{} design.

Following the design of Megatron-LM~\cite{shoeybi2019megatron} in its tensor parallel(TP) and data parallel(DP) arrangement, Pier first ensures that TP ranks are placed within the same node whenever possible. 
During the lazy start phase, \name{} uses AdamW in data parallel over tensor parallel, as shown in Figure~\ref{fig:tp_figure}.
In this example, a model is split across two processors in every data parallel rank.
Each data parallel rank (DP rank) maintains a complete copy of the model, while the model is split across two tensor parallel ranks (TP ranks). 
So all processors with the same TP rank have a synchronized copy of the model partition.
And they can compute the momentum by comparing the changes of the current value and the previous value.
The warmup momentum for the first model partition is available on all processors with TP0.
And the momentum for the second model partition is available on all processors with TP1. 

Upon transition, the inner optimizer (i.e., AdamW) communicates the gradients of DP0 and DP1 in Node 0.
A concurrent all-reduce operation runs on DP2 and DP3 in Node 1. 
All the communication of inner optimizers are within a node, thus it can benefit from the high-bandwidth links, e.g., NVLink. 
The outer optimizer is activated periodically, e.g., for every 50 iterations. 
Upon activation, the outer optimizer will all-gather the model partitions among the processors with the same TP rank. 
In this case, the all-gather is on four processors.
This communication spans multiple nodes over low-bandwidth links, such as InfiniBand.
However, global communication is only required every 50 iterations. 
As shown in Figure~\ref{fig:tp_figure}, the all-gather on TP ranks of 0 does not overlap with TP ranks of 1.
The communication can be launched in parallel naturally, which allows for efficient utilization of the low-bandwidth links. 

Although \name{} only implements DP and TP, it can be easily extended to include pipeline parallelism, where the all-gather of model states can be streamlined.
The current implementation is compatible with FSDP-1 (fully sharded data parallel) and FSDP-2, as FSDP-1 only distributes optimizer states over processors. 
Additionally, the extra model distribution strategy in FSDP-2 can be integrated in the same manner as TP.

In this paper, we demonstrate the applicability of \name{} in DP and TP, highlighting the runtime speedup achieved by \name{} compared to AdamW.

%So that every processor has a synchronized copy of the model.
%Each processor can compute the momentum by comparing the changes of the current model and previous model.
%The warmup momentum is available on every processor with identical values. 

%the warmup moemntum is distributed accross tensor-parallel ranks. 
%In the subsequent low-communication stage, Pier further divides each groups into smaller sub-communication groups. Within these groups, inner gradient synchronization occurs locally in non-synchronization iterations, and outer synchronization happens at periodic intervals. Both inner and outer gradient synchronizations, whether it's full DP or DiLoCo, are performed within each TP rank and executed in parallel. 
\section{Implementation}

\label{sec:impl}
We prototype \name{} on top of Megatron-LM~\cite{megatron-github}.
The local communication data parallel groups (referred to as groups in the following) are formed by partitioning the data parallel ranks during the communication initialization process.
\name{} uses AdamW as the inner optimizer within each group and Nestorov momentum as the outer optimizer across the groups.
The outer synchronization (i.e., model gathering) is integrated into the main training loop by the end of each interval.
To fully utilize the compute capability, \name{} leverages FlashAttention-2~\cite{dao2023flashattention}.
%The momentum is stored in BF16 with the native support of PyTorch.

Our selection of outer optimizer follows the conclusion of the DiLoCo paper~\cite{douillard2023diloco}, which states that the Nesterov momentum optimizer achieves the best results. 
However, it is notable that there is a difference between the theoretical Nesterov momentum and the PyTorch implementation. 
In the original Nesterov momentum~\cite{nesterov1983method}, the update is performed by first taking a look-ahead step in the direction of the current momentum, then computing the gradient at this anticipated position, and finally correcting the momentum and model weights based on this look-ahead gradient. 
However, the PyTorch implementation of Nesterov momentum ~\cite{pytorch-sgd-doc} uses an approximated formulation, where the direction of update is efficiently adjusted by combining the momentum and the current gradient in one step, without explicitly performing the separate look-ahead gradient evaluation. 
We implement and test both versions, and select the PyTorch version of Nesterov momentum since it achieves better empirical performance in our setting. 

The original DiLoCo work~\cite{douillard2023diloco} suggests a fixed outer learning rate because the outer gradient is proportional to the inner update, which is already controlled by the inner learning rate scheduler. 
However, our empirical results show that adjusting the outer learning rate improves both the stability in the early stage and the final convergence. 
We introduce an outer learning rate warmup phase for the first 10\% to 20\% of the training, where the outer learning rate linearly increases from 0 to 1. 
This warmup phase stabilizes the training in the early phase. 
After the warmup, the learning rate is increased to 1.1 between 20\% and 80\% for faster convergence; then we reduce it to 0.9 for the final 20\%, since the recommended value from DiLoCo paper is 0.7, which is relatively small compared to our choices. 
% matching the recommended value from DiLoCo paper. 

The outer optimizer in \name{} evaluates the gradients by differencing the models' states.
Thus, it needs to store an additional version of the model, which exacerbates the already intensive memory usage.
To lower the memory consumption of the outer optimizer, \name{} offloads the old model to host memory.
At the beginning of the outer optimizer steps, \name{} reloads the old model states, gathers the latest model states, and evaluates the gradients.
At the end of the outer optimizer steps, \name{} will offload the latest model states to host memory and release the memory in GPUs.
Under standard DP, on a compute node with multiple GPUs, each GPU partitions the model and only offloads and reloads the corresponding model shard and the outer optimizer momentum.
This is to avoid redundant data movement between host memory and GPU memory. 
Also, Pier offloads the momentum of the outer optimizer to host memory with a similar approach to further reduce the memory footprint. 
We implement a switch to enable/disable CPU offload, as offload introduces a trade-off between I/O overhead and memory footprint.

%We also implemented an CPU offload mechanism to reduce GPU memory usage. During low-communication training, the outer optimizer needs to access the model parameter from the last synchronization point, which could take a lot of memory. Pier offloads the model parameters to CPU memory at the end of each outer step, and reloads this model parameter when it's needed, that is, at the beginning of the next outer step. To be specific, each GPU with DP rank 0 loads the model shard and then broadcasts the parameters across their global data parallel group, which avoids excessive concurrent CPU I/O and take advantage of the high-bandwidth GPU interconnect.

\section{Experiments}
\label{sec:expr}
In this section, we study the convergence and scalability of \name{}.
Across all experiments, we pretrain GPT-2 small, medium, XL, and the 7B models on in OpenWebText dataset using mixed-precision (BF16 in models and FP32 in optimizers) from Megatron-LM. 
The hyperparameters of the experiments are in Table~\ref{tb:setting}.

\begin{table}[ht]
\caption{Hyperparameter Setting. We followed the hyperparameter used in Sophia~\cite{liu2023sophia}.}
\label{tb:setting}
\vskip 0.15in
\begin{center}
\begin{scriptsize}
\begin{sc}
\begin{tabular}{lcc}
\toprule
\textbf{Hyperparameter} & Value\\
\midrule
inner optimizer & AdamW \\
inner learning rate (GPT-2 small, medium, XL) & 4e-4, 3e-4, 1.5e-4 \\
inner minimum learning rate & 4e-5, 3e-5, 1.5e-5 \\
adam beta1 & 0.9 \\
adam beta2 & 0.999 \\
inner decay style & cosine\\
global batch size & 512 \\
Training iterations & 100000 \\ 
inner decay iterations & 100000\\
learning rate warmup proportion & 2\% \\ 
weight decay  & 0.1 \\
clip grad & 1\\

\midrule
outer optimizer & Nesterov\\
synchronization interval \textit{H} & 50 ,100, 200, 500\\
outer learning rate & in Section~\ref{sec:impl}\\
outer momentum coefficient & in Section~\ref{sec:design:decay} \\
number of groups verified for convergence& 8, 32, 64 \\
\bottomrule
\end{tabular}
\end{sc}
\end{scriptsize}
\end{center}
\vskip -0.1in
\end{table}
%\zhao{add this table, and say that the LR, steps follow other research, such as Sophia.}

\begin{figure*}[htb]
  \centering
  \subfigure[GPT-2 small]{
    \includegraphics[width=0.31\textwidth]{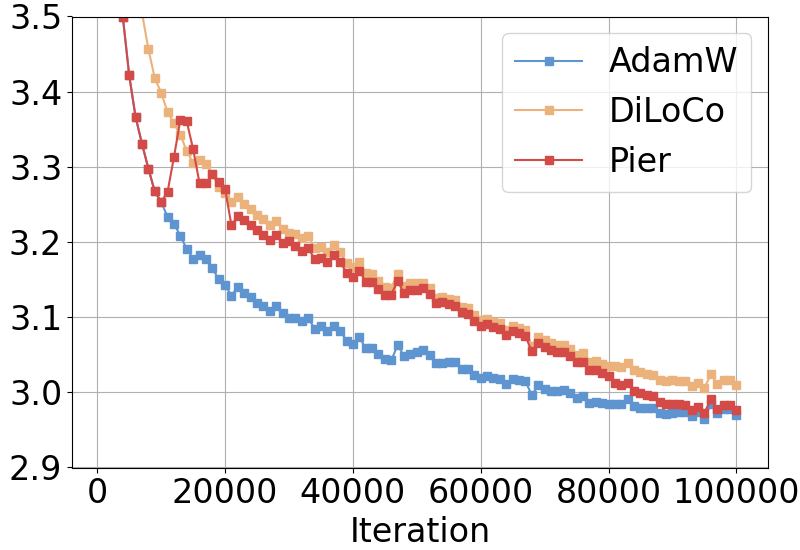}
    \label{fig:scale}
  }
  \hfill
    \subfigure[GPT-2 medium]{
    \includegraphics[width=0.31\textwidth]{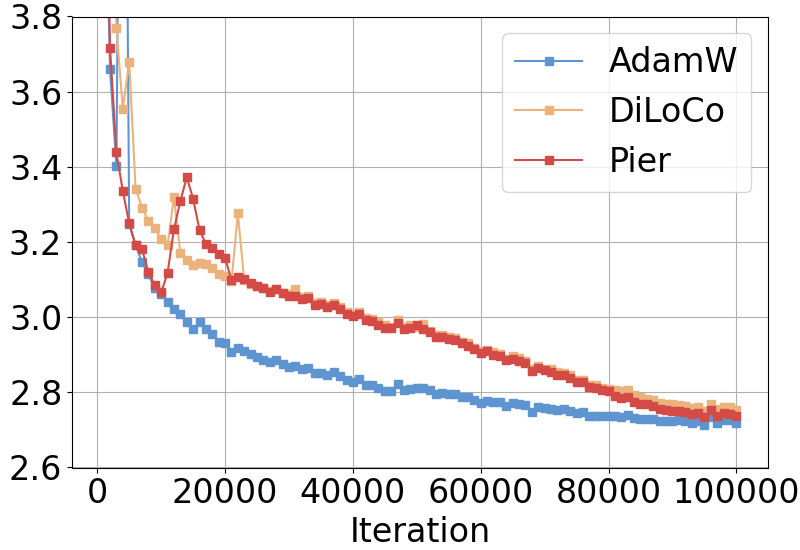}
    \label{fig:scale}
  }
  \hfill
    \subfigure[GPT-2 XL]{
    \includegraphics[width=0.31\textwidth]{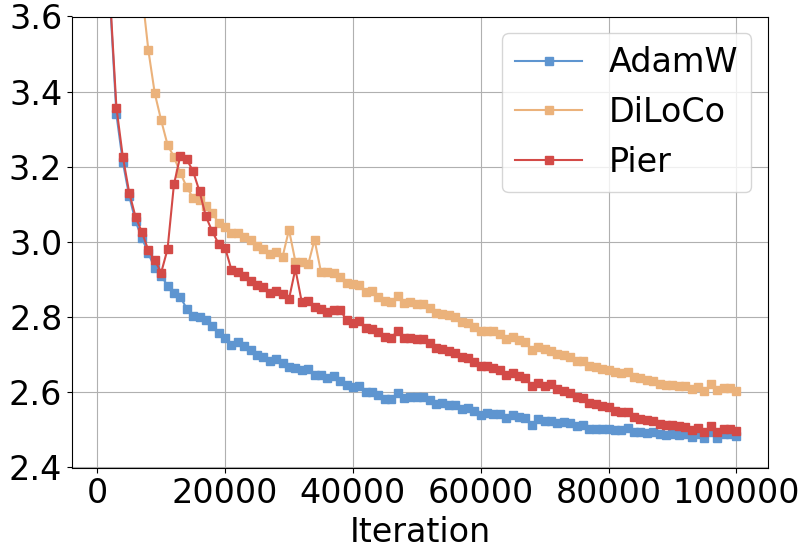}
    \label{fig:scale}
  }
  \hfill
  
  %\vspace{0.1in}
  \caption{\fontsize{9pt}{11pt}\selectfont
  Validation Loss Curves of GPT-2 Small, Medium and XL during Pretraining with AdamW, DiLoCo, and Pier. For GPT-2 small and XL, our approach achieves validation losses that are closer to those of AdamW. For GPT-2 Medium, our approach achieves lower validation loss compared to the original DiLoCo.}
  \label{fig:default_convergence}
  %\vspace{0.1in}
\end{figure*}

\begin{table*}[t]
\caption{Downstream Tasks of GPT-2 Small (S), Medium (M), XL Models Pretrained with AdamW, DiLoCO, and \name{}. Mul-rc is MultiRC; Rcd is ReCorD; M-qa is MathQA; Lamb. is LAMBADA-OPENAI; Wino. is Winograd.}
\label{tab:downstream-XL}
\vskip 0.15in
\begin{center}
\begin{scriptsize}
\begin{sc}
\begin{tabular}{lccccccccccccccc}
\toprule
\textbf{Method} & boolq & cb-acc & copa& mul-rc& rcd-f1& rte& wic&wsc&lamb.&race&m-qa&piqa&wino.\\
\midrule

AdamW-S & \textbf{0.5272} & 0.3750 & 0.6200 & 0.5474 & 0.7157 & \textbf{0.5090} & 0.5000 & 0.3558 & \textbf{0.3324} & 0.2852 & \textbf{0.2174} & \textbf{0.6148} & 0.4996  \\
DiLoCo-S & 0.5098 & 0.1964 & \textbf{0.6500} & \textbf{0.5703} & 0.6976 & 0.4838 & \textbf{0.5031} & \textbf{0.3654} & 0.3167 & 0.2804 & 0.2107 & 0.6023 & 0.5130  \\
Pier-S & 0.4963 & \textbf{0.4821} & 0.6400 & \textbf{0.5703} & \textbf{0.7182} & 0.5054 & 0.4953 & \textbf{0.3654} & 0.3171 & \textbf{0.2957} & 0.2141 & \textbf{0.6148} & \textbf{0.5146} \\
\midrule
AdamW-M & 0.5560 & 0.3750 & 0.6900 & \textbf{0.5458} & 0.7873 & 0.5307 & 0.4906 & 0.3654 & 0.4186 & 0.2986 & 0.2258 & \textbf{0.6540} & 0.5462  \\
DiLoCo-M & 0.5804 & 0.3393 & \textbf{0.7000} & 0.5336 & 0.7907 & 0.5343 & \textbf{0.5000} & \textbf{0.3750} & \textbf{0.4277} & \textbf{0.3053} & \textbf{0.2271} & 0.6431 & 0.5288  \\
Pier-M & \textbf{0.5872} & \textbf{0.5357} & \textbf{0.7000} & 0.5264 & \textbf{0.7953} & \textbf{0.5668} & \textbf{0.5000} & 0.3654 & 0.4120 & 0.2967 & 0.2181 & 0.6485 & \textbf{0.5478}  \\
\midrule
AdamW-XL & 0.5942 & \textbf{0.5179} & \textbf{0.7400} & 0.4767 & 0.8445 & 0.4946 & \textbf{0.5016} & 0.4038 & 0.4995 & 0.3330 & 0.2332 & 0.6899 & 0.5541  \\
DiLoCo-XL & \textbf{0.6128} & 0.1607 & 0.6600 & \textbf{0.4878} & 0.8285 & 0.5596 & 0.5000 & 0.3654 & 0.4733 & 0.3225 & 0.2275 & 0.6746 & 0.5612  \\
Pier-XL & 0.5174 & 0.2500 & \textbf{0.7400} & 0.4775 & \textbf{0.8532} & \textbf{0.5632} & 0.4828 & \textbf{0.6346} & \textbf{0.5331} & \textbf{0.3502} & \textbf{0.2379} & \textbf{0.6986} & \textbf{0.5888} \\
\bottomrule
\end{tabular}
\end{sc}
\end{scriptsize}
\end{center}
\vskip -0.1in
\end{table*}

\subsection{Convergence and Effectiveness Boundary}
For the \textbf{convergence} study, we use the metrics of validation loss and a suite of thirteen downstream tasks.
This task suite comprises eight tasks from the SuperGLUE benchmark~\cite{wang2019superglue}, along with five additional tasks: LAMBADA, RACE, MathQA, PIQA, and Winograd.  
Throughout this study, we use two baselines: AdamW in standard data parallelism and the original DiLoCo.
The convergence result is in \S\ref{sec:expr:convergence}.
For \name{}, the convergence depends on the global batch size and the synchronization interval. 
Thus, we perform two experiments to quantify the effectiveness boundary of the two factors.
The results are reported in \S\ref{sec:expr:weak_convergence} and \S\ref{sec:expr:interval_convergence}, respectively.

\subsubsection{Convergence}
\label{sec:expr:convergence}
The first experiment is to verify the convergence of \name{}.
Thus, we run full pretraining (i.e., 100,000 iterations) with AdamW, DiLoCo, and \name{}.
We measure the convergence using validation loss and the thirteen downstream tasks.

Figure~\ref{fig:default_convergence} presents the validation loss curves on GPT-2 small, medium, and XL with 8, 32, 64 groups, with one GPU in each group.
For \name{}, we observe a clear switching point, as the model suffers from a sudden loss spike.
\name{} successfully mitigates the loss instability with the momentum warmup and decay techniques, showing faster convergence compared to the original DiLoCo.
For GPT-2 small, the validation loss of AdamW is 2.97, compared to 3.00 of DiLoCo and 2.97 of \name{}.
For GPT-2 medium, the validation losses are 2.72, 2.75, and 2.74 for AdamW, DiLoCo, and \name{}, respectively.
For GPT-2 XL, AdamW achieves a validation loss of 2.48.
The loss value of DiLoCo is 2.60 and \name{}'s value is 2.49. 
\name{} consistently achieves lower validation loss than the original DiLoCo, and the loss is comparable to the AdamW baseline.
Table~\ref{tab:downstream-XL} shows the downstream task performance across the three optimizers.
For GPT-2 small, AdamW has five tasks with the best performance, while DiLoCo and \name{} each have four and seven.
For GPT-2 XL, \name{} has nine tasks with the best performance, compared to three for AdamW and two for DiLoCo.

These results confirm that the convergence of \name{} is comparable to AdamW.
Across the three models, \name{} consistently shows better downstream task performance and lower validation loss than the original DiLoCo.

%The first thing we want to verify is that our approach can converge properly. Thus, we run full pretraining with three candidate approaches. We measure convergence by evaluating validation loss during pretraining, and downstream task performance after pretraining. Figure~\ref{fig:default_convergence} presents the validation loss curves on GPT-2 small, medium and XL with 8, 32, 64 groups. Compared with original DiLoCo, our method exhibits a more consistent result with the standard data parallelism. At the switching point, the model suffers from a sudden loss increase for a while, but our \name{} momentum with momentum decay technique could successfully handle this loss instability, achieving a faster convergence compared with original diloco. This figure shows the key contribution of our \name{} approach: stabilize the switching point and taking advantage of the fast early convergence of full data parallel. Table~\ref{tab:downstream-XL} also shows the downstream task of final model, including GPT-2 small, medium, XL, pre-trained with standard data parallelism, original DiLoCo and \name{}. Our models have better scores, which are more close to the AdamW baseline. These results shows Pier has no model performance degradation compared to the standard AdamW approahch.

\subsubsection{Global Batch Size}
\label{sec:expr:weak_convergence}
In this experiment, we explore the global batch size boundary of \name{}, as researchers can train the same model with more GPUs if model performance is guaranteed with larger batch sizes.
We perform this experiment by pretraining GPT-2 small using weak scaling, where the global batch size increases in proportion to the GPU size.
So that each GPU can maintain the same workload to sustain high utilization.
To preserve fairness, we maintain a fixed budget of tokens across  all scales. 
E.g., the number of iterations is halved with a doubled global batch size, i.e., the GPU count.
So the global batch sizes on the four scales of \{4, 8, 16, 32\} GPUs are 256, 512, 1024, and 2048.
And the numbers of iterations are 200,000, 100,000, 50,000, and 25,000.

\begin{figure}[ht]
  \centering
  \includegraphics[width=\columnwidth]{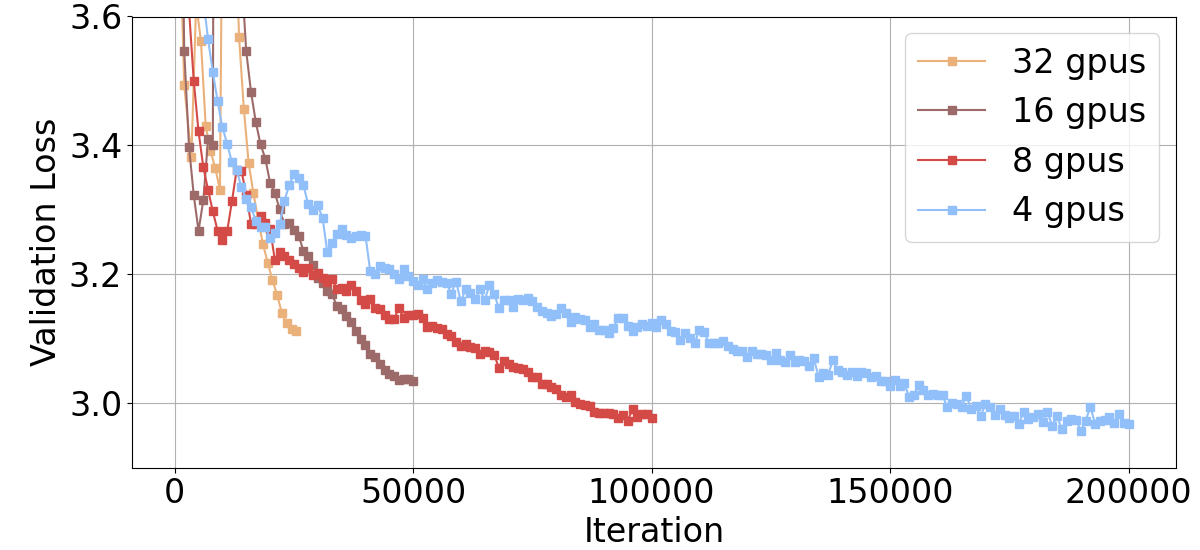}
  %\vspace{0.1in}
  \caption{\fontsize{9pt}{11pt}\selectfont
  Validation Loss Curves of Pier under Weak Scaling. The runs with 4 GPUs exhibit convergence. When scaling up to 16 or 32 GPUs, the validation loss rises significantly compared with the default 8 GPUs setting.}
  \label{fig:weak_curve}
  %\vspace{0.1in}
\end{figure}

The validation curves are shown in Figure~\ref{fig:weak_curve}. 
From four, eight, sixteen, to thirty-two GPUs, the validation loss increases monotonically with the values of 2.96, 2.97, 3.03, and 3.10.
Table~\ref{tb:weak} presents the downstream task performance across GPU counts. 
The model performance decreases with an increasing number of GPUs.
With four GPUs (i.e., global batch size of 256), \name{} has nine tasks with equal or higher performance than the AdamW baseline.
This number decreases to eight on eight GPUs, seven on sixteen GPUs, and six on thirty-two GPUs. 

The results confirm that the effective global batch size boundary for \name{} is 512 for GPT-2 small with the model performance guarantee.
We assume the same global batch size boundary for other models in the subsequent study. 
This is a conservative assumption, with which our subsequent study is guaranteed to converge.

%In pretraining context, weak scaling refers to increasing the number of computing devices while proportsionally increasing the global batch size so that the workload per device remains unchanged. In this section we aims to explore the boundary of weak scaling while maintaining model convergence. In our setting, we increase the global batch size and reduce the number of iterations to keep the total number of tokens processed constant in pretraining of GPT-2 small. Our scaling starts from 8 GPU, whose hyper-parameter settings remains identical to default convergence experiment. For 4 GPUs we halve the global batch size, double the number of steps with 4 groups. For 16 GPUs we double the batch size and halve the number of steps with 16 groups, and for 32 GPUs we scale the batch size by 4 times while reducing the iteration to $\frac{1}{4}$ with 32 groups. The validation loss curves are shown in Figure~\ref{fig:weak_curve}, and downstream tasks results are shown in Table~\ref{tb:weak}. Our results, including the valiation loss and downstream tasks, clearly shows that weak scaling down to 4 GPUs won't introduce errors, but when weakly scaling up to 16 or 32 GPUs, there is a slightly model performance degradation with a significantly validation loss increase.

\begin{table*}[tb]
\caption{Downstream Tasks Results of GPT-2 Small Models Pretrained with Different Number of GPUs in Weak Scaling. Mul-rc is MultiRC; Rcd is ReCorD; M-qa is MathQA; Lamb. is LAMBADA-OPENAI; Wino. is Winograd; Loss is validation loss. Bold values are equal or higher performance than the AdamW baseline in Table~\ref{tab:downstream-XL}.}
\label{tb:weak}
\vskip 0.15in
\begin{center}
\begin{scriptsize}
\begin{sc}
\begin{tabular}{lcccccccccccccccc}
\toprule
\textbf{GPU} & boolq & cb-acc & copa& mul-rc& rcd-f1& rte& wic&wsc&lamb.&race&m-qa&piqa&wino.&loss\\
\midrule

%AdamW & 0.5272 & 0.3750 & 0.6200 & 0.5474 & 0.7157 & 0.5090 & 0.5000 & 0.3558 & \textbf{0.3324} & 0.2852 & 0.2174 & \textbf{0.6148} & 0.4996 & 2.97 \\
4 & \textbf{0.5355} & \textbf{0.4643} & \textbf{0.6700} & \textbf{0.5631} & \textbf{0.7207} & \textbf{0.5235} & \textbf{0.5157} & 0.3462 & 0.3315 & \textbf{0.2967} & 0.2141 & 0.6094 & \textbf{0.5051} & \textbf{2.96}\\
8 & 0.4963 & \textbf{0.4821} & \textbf{0.6400} & \textbf{0.5703} & \textbf{0.7182} & 0.5054 & 0.4953 & \textbf{0.3654} & 0.3171 & \textbf{0.2957} & 0.2141 & \textbf{0.6148} & \textbf{0.5146} & 2.97 \\
16 & 0.4514 & 0.2679 & \textbf{0.6800} & \textbf{0.5501} & 0.7030 & \textbf{0.5307} & \textbf{0.5078} & \textbf{0.3654} & 0.3136 & 0.2823 & \textbf{0.2228} & 0.6001 & \textbf{0.5075}  & 3.03 \\
32 & 0.5153 & 0.3214 & \textbf{0.6500} & 0.5311 & 0.6749 & \textbf{0.5343} & \textbf{0.5000} & \textbf{0.3654} & 0.3022 & \textbf{0.2890} & 0.2097 & 0.6023 & \textbf{0.5178} & 3.10 \\
\bottomrule
\end{tabular}
\end{sc}
\end{scriptsize}
\end{center}
\vskip -0.1in
\end{table*}

\subsubsection{Synchronization Interval}
\label{sec:expr:interval_convergence}
To understand the impact of the synchronization interval in the outer optimizer on convergence, we vary the interval with \{50, 100, 200, 500\} iterations on GPT-2 small pretraining.
We then compare the validation loss and downstream task performance derived from each interval against the AdamW baseline (see Table~\ref{tab:downstream-XL}).

Table~\ref{tb:sync} shows the model performance of different synchronization intervals. 
All four intervals show comparable validation loss values of around 2.97 at the end of the 100,000-iteration pretraining.
For downstream tasks, the intervals of \{50, 100, 200, 500\} have \{8, 4, 9, 8\} tasks with equal or higher performance than the AdamW baseline.
There is no clear pattern in the model performance across the interval range.

The results show that the convergence of \name{} is less sensitive to synchronization intervals, even for extremely large intervals, such as 500 iterations.

%Despite the large effective interval of 500 iterations, we decide to choose the 50 iterations as the default interval in the speedup measurement study in \S\ref{sec:expr:performance_strong} and \S\ref{sec:expr:number_groups} as a conservative decision.

%Although the DiLoCo paper ~\cite{douillard2023diloco} has demonstrated a validation loss function difference between synchronization intervals, we further examine the impact of increasing synchronization interval on convergence, under our Pier setting with full downstream tasks evaluations. Table~\ref{tb:sync} shows the model performance of different synchronization intervals. Our results shows that Pier has a solid convergence with increased synchronization invervals, even at extremely large intervals like 500 steps.
\begin{table*}[tb]
\caption{Downstream Task Results of GPT-2 Small Models Pretrained with Different Synchronization Intervals. Int. is synchronization interval; Mul-rc is MultiRC; Rcd is ReCorD; M-qa is MathQA; Lamb. is LAMBADA-OPENAI; Wino. is Winograd; Loss is validation loss. Bold values are equal or higher performance than the AdamW baseline in Table\ref{tab:downstream-XL}. }
\label{tb:sync}
\vskip 0.15in
\begin{center}
\begin{scriptsize}
\begin{sc}
\begin{tabular}{lcccccccccccccccc}
\toprule
\textbf{Int.} & boolq & cb-acc & copa& mul-rc& rcd-f1& rte& wic&wsc&lamb&race&math&piqa&wino.&loss\\
\midrule

%AW. & \textbf{0.5272} & 0.3750 & 0.6200 & 0.5474 & 0.7157 & \textbf{0.5090} & \textbf{0.5000} & 0.3558 & \textbf{0.3324} & 0.2852 & 0.2174 & 0.6148 & 0.4996 & 2.97 \\
50 & 0.4963 & \textbf{0.4821} & \textbf{0.6400} & \textbf{0.5703} & \textbf{0.7182} & 0.5054 & 0.4953 & \textbf{0.3654} & 0.3171 & \textbf{0.2957} & 0.2141 & \textbf{0.6148} & \textbf{0.5146} & 2.97 \\
100& 0.4211 & 0.3571 & \textbf{0.6500} & 0.5221 & 0.7142 & 0.5018 & 0.4734 & \textbf{0.3846} & 0.3138 & \textbf{0.2919} & 0.2124 & 0.6083 & \textbf{0.5051} & 2.97\\

200&0.5125 & \textbf{0.5357} & \textbf{0.6200} & \textbf{0.5633} & \textbf{0.7170} & 0.4585 & \textbf{0.5000} & \textbf{0.3654} & 0.3266 & \textbf{0.3072} & 0.2157 & \textbf{0.6148} & \textbf{0.5020} & 2.97 \\
500&0.4453 & \textbf{0.4107} & \textbf{0.6200} & \textbf{0.5660} & \textbf{0.7202} & 0.4910 & 0.4922 & \textbf{0.3942} & 0.3115 & 0.2813 & \textbf{0.2181} & \textbf{0.6192} & \textbf{0.5162} & 2.98\\
\bottomrule
\end{tabular}
\end{sc}
\end{scriptsize}
\end{center}
\vskip -0.1in
\end{table*}

\subsection{Runtime Speedup}
Upon confirmation of convergence, we investigate the runtime speedup patterns of \name{} under various settings.
In \S\ref{sec:expr:performance_strong}, we measure the runtime speedup of \name{} compared to AdamW with a fixed number of groups and global batch size.
In \S\ref{sec:expr:number_groups}, we vary the number of groups while keeping the group size constant on the two GPU clusters.
We report how \name{} performs with data parallel and tensor parallel for LLM pretraining in \S\ref{sec:expr:TP}.

% \zhao{Define scaling efficiency, speedup, and performance improvement.}
To quantify runtime improvements, we evaluate three metrics. First, speedup $S$ is the runtime ratio of the AdamW baseline ($T_{baseline}$) to our proposed method \name{} (${T_{Pier}}$):
$$S = \frac{T_{baseline}}{T_{Pier}}$$
Second, performance improvement $\Delta p $ is the corresponding percentage reduction in runtime:
$$\Delta p = \frac{T_{baseline}-T_{Pier}}{T_{baseline}}\times 100\%$$
Finally, scaling efficiency $e$ measures how well the runtime decreases for a fixed problem size as the number of processing elements increases from M to N. It's defined as: 
$$e = \frac{T_{M}}{T_{N}}\times \frac{M}{N}$$
The reference value $M$ depends on the experimental setup: in Section~\ref{sec:expr:performance_strong}, $M$ is set to 8, 32, and 64 for the GPT-2 small, medium, and XL models, respectively; while in Section~\ref{sec:expr:number_groups}, $M$ is set to 1.
\begin{figure*}[ht]
  \centering
  \subfigure[GPT-2 small]{
    \includegraphics[width=0.31\textwidth]{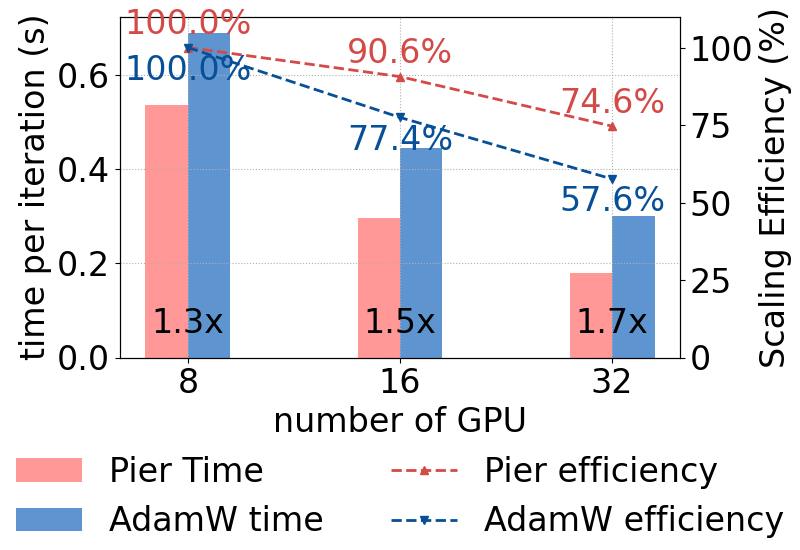}
    \label{fig:s-s}
  }
  \hfill
    \subfigure[GPT-2 medium]{
    \includegraphics[width=0.31\textwidth]{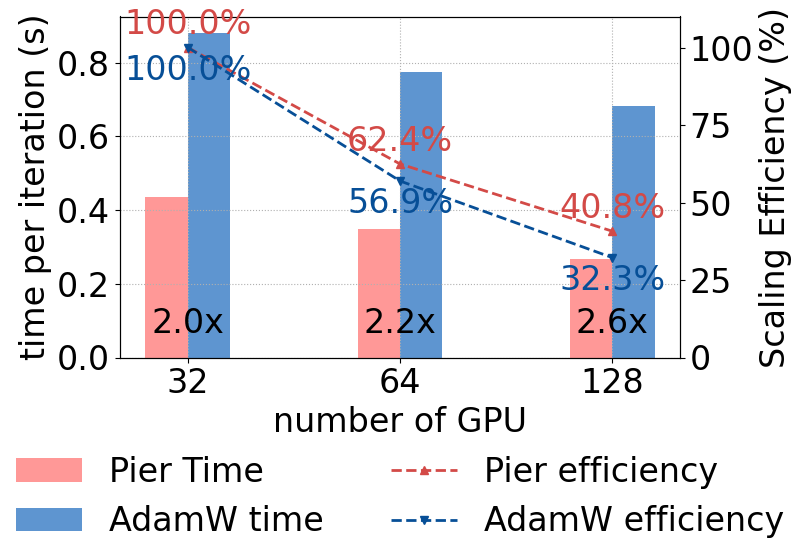}
    \label{fig:s-m}
  }
  \hfill
    \subfigure[GPT-2 XL]{
    \includegraphics[width=0.31\textwidth]{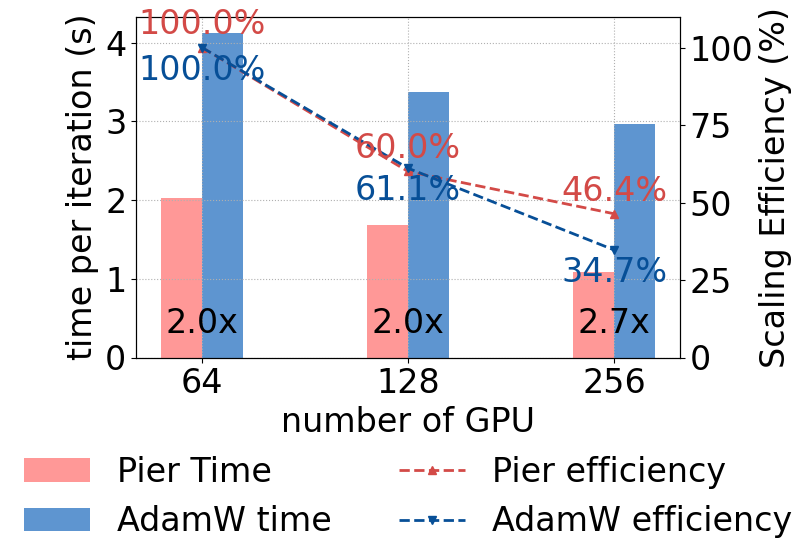}
    \label{fig:s-x}
  }
  \hfill
  
  %\vspace{0.1in}
  \caption{\fontsize{9pt}{11pt}\selectfont
  Runtime and Strong Scaling Efficiency Comparison Between AdamW and Pier. The proposed method achieves up to 1.7x, 2.6x, and 2.7x speedup for pretraining of GPT-2 small, medium, and XL on the NERSC Perlmutter Cluster. Number of communication groups is set to \{8, 32, 64\} in Figure~\ref{fig:s-s}, ~\ref{fig:s-m},~\ref{fig:s-x}, which has a verified convergence in \ref{sec:expr:convergence}.}
  \label{fig:strong}
  %\vspace{0.1in}
\end{figure*}
\subsubsection{Strong Scaling}
\label{sec:expr:performance_strong}
The previous experiments have validated the convergence property of \name{} as well as the effectiveness boundary of global batch size, number of groups, and the synchronization interval in the outer optimizer.
In this study, we profile the runtime speedup of \name{} compared to AdamW in standard data parallelism.

The experiments run on NERSC Perlmutter, where each compute node is equipped with four NVIDIA A100 GPUs.
We fix the synchronization interval at 50 iterations to estimate the lower bound of the runtime improvement.
We also project the runtime reduction with an interval of 500 iterations as the upper bound estimation.
The number of groups is also fixed with the empirical settings that have shown convergence.
For example, \name{} uses 64 groups (one GPU per group) in the convergence study in \S\ref{sec:expr:convergence} for GPT-2 XL.
Then we increase the group size (i.e., the number of GPUs in a group) from one, two, to four GPUs, to limit intra-group communication within a node.

Figure~\ref{fig:strong} shows the runtime measurement and scaling efficiency of GPT-2 models with \name{} and AdamW.
We run each experiment for 2,000 iterations, then estimate the speedup by projecting to the full 100,000-step pretraining.
Please note that we set the global batch size to 512.
On smaller scales, the local batch size per GPU is eight. 
Megatron-LM accumulates gradients to realize a larger batch size.
On larger scales, such as 128 GPUs, we must lower the local batch size to four, which may result in lower utilization on each GPU, as the batch size may not saturate the GPU's computing capability.
The comparison between \name{} and AdamW is still fair, as the local batch size per GPU is identical in both cases.
The time estimation is calculated by a weighted average, with 10\% momentum warmup using AdamW in standard data parallel and 90\% with \name{}. 

\begin{figure}[ht]
  \centering
  \includegraphics[width=1\columnwidth]{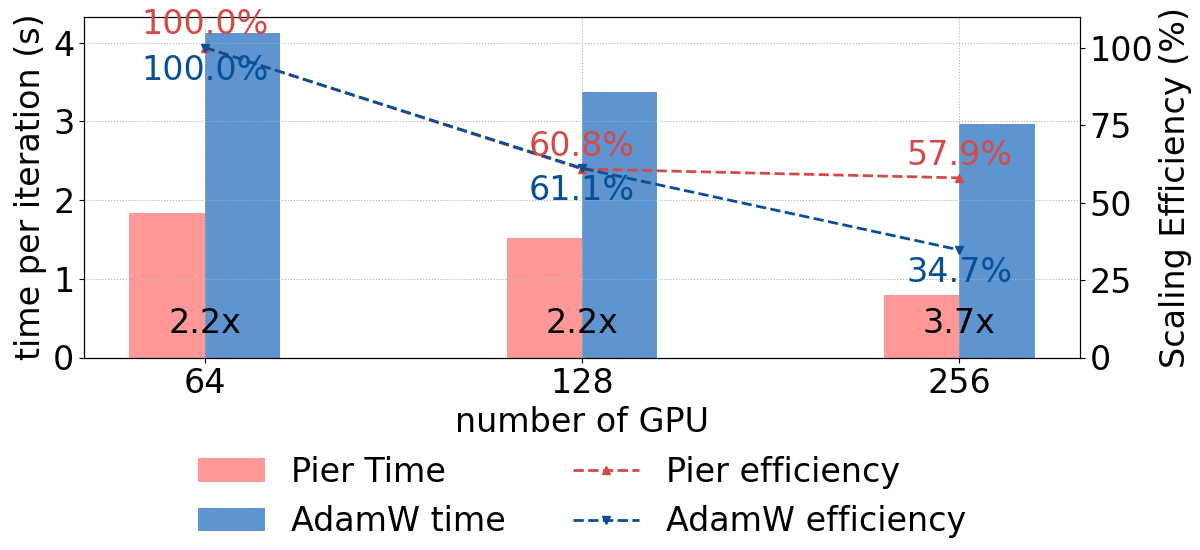}
  %\vspace{0.1in}
  \caption{\fontsize{9pt}{11pt}\selectfont
  Runtime and Strong Scaling Efficiency Comparison between AdamW and Pier. The experiment follows the same basic setting as in Figure~\ref{fig:s-x}, but Pier achieves better results with a synchronization interval of 500.}
  \label{fig:predict500}
  %\vspace{0.1in}
\end{figure}

For GPT-2 models, \name{} speeds up pretraining by 1.7x, 2.6x, and 2.7x for small, medium, and XL, respectively, on the largest possible scale.
With the synchronization interval of 500 iterations, the speedup of \name{} on GPT-2 XL is 2.2x, 2.2x, and 3.7x on 64, 128, and 256 A100s, as shown in Figure~\ref{fig:predict500}.
The sustained scaling efficiency is 57.9\% compared to 34.7\% for AdamW on 256 A100s.

\begin{figure*}[ht]
  \centering
  \subfigure[Perlmutter]{
    \includegraphics[width=0.48\textwidth,trim=12mm 0mm 0mm 0mm, clip]{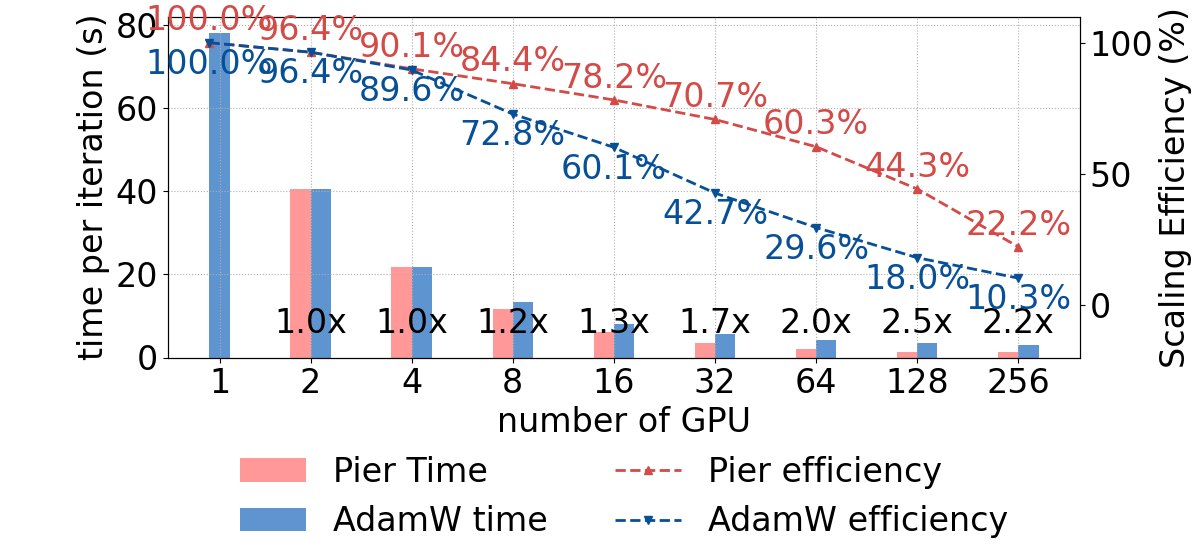}
    \label{fig:scale-p}
  }
  \subfigure[Vista]{
    \includegraphics[width=0.48\textwidth,trim=12mm 0mm 0mm 0mm, clip]{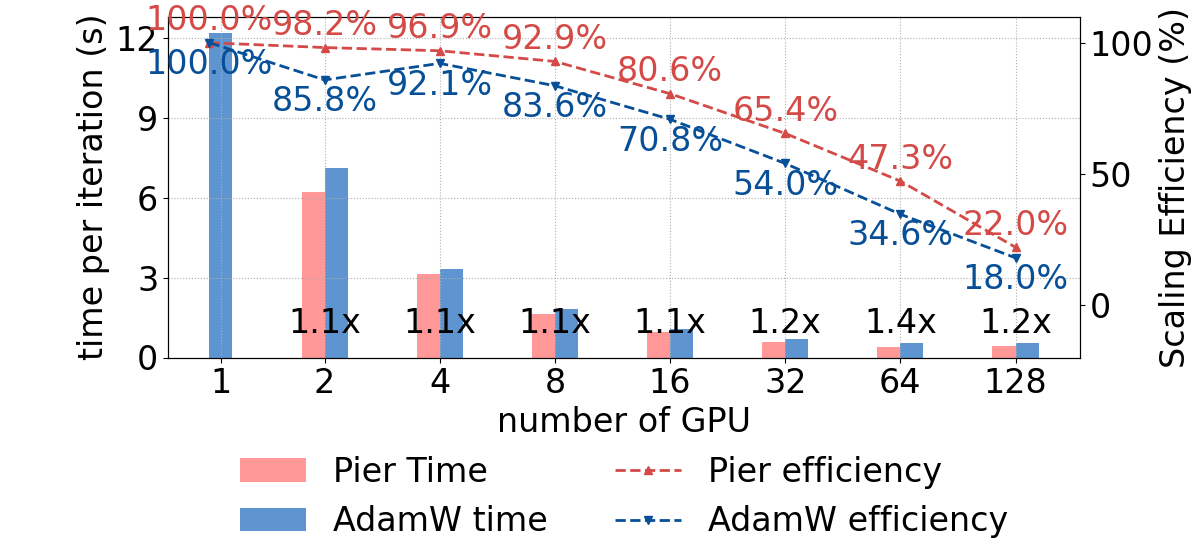}
    \label{fig:scale-v}
  }
  %\vspace{0.1in}
  \caption{\fontsize{9pt}{11pt}\selectfont
    Runtime and Scaling Efficiency Comparison between AdamW and Pier of GPT-2 XL Pretraining, where the number of communication groups equals to the number of GPUs. The proposed method achieves up to 2.5x speedup using 128 A100 GPUs on Perlmutter, and up to 1.4x speedup when using 64 GH200 GPUs on Vista. The scaling efficiency is calculated based on the runtime of AdamW with 1 GPU.}
  \label{fig:scale-dual}
  %\vspace{0.1in}
\end{figure*}

\subsubsection{Performance: Varying number of groups}
\label{sec:expr:number_groups}
In this experiment, we examine the runtime performance gain when each GPU is considered as a group.
This setting completely eliminates the communication in the inner optimizer. 
We fix the global batch size across scales for convergence guarantee.
We measure the speedup on both NERSC Perlmutter and TACC Vista.

Figure~\ref{fig:scale-dual} shows the measurements of GPT-2 XL.
On Perlmutter, \name{} achieves significant speedups beyond one compute node (i.e., four GPUs).
The speedup increases monotonically from 4 to 128 GPUs, then decreases at 2.2x on 256 A100s.
The scaling efficiency is 70.7\%, which is drastically higher than the 42.7\% achieved by AdamW at the same scale. 
On Vista, \name{} is more effective at larger scales. 
E.g., \name{} shows a speedup of 1.4x and 1.2x on 64 GH200s and 128 GH200s, respectively.

With a further relaxed communication interval of 500 iterations, the speedup in runtime is 1.8x and 1.9x for \name{} on 64 and 128 GH200s, respectively.
The relatively lower speedup on Vista compared to Perlmuter can be attributed to the machine configuration.
On Vista, each node is equipped with one GH200 superchip, and the network is Infiniband NDR, which offers a bandwidth of 400~Gbps.
The network is shared with 256 CPU nodes and 600 GPU nodes.
In contrast, the compute node on Perlmutter has four A100 GPUs connected by NVLink, and the network is Slingshot 11 with four NICs per node.
The network is shared by 3,072 CPU nodes and 1,792 GPU nodes.
The scaling efficiency of GPT-2 XL on 64 GH200 GPUs is reported as 34.6\% for AdamW. 
\name{}'s scaling efficiency is 47.3\% on the same scale,  which is 36.7\% higher than AdamW

With the synchronization interval of 50 iterations, the scaling efficiency of \name{} drops below 40\% beyond 64 GH200s and 128 A100s, which are the efficient scaling boundaries for GPT-2 XL. 
The boundary depends on the model, the synchronization interval, the computing capability of GPUs, and the networks.

\subsubsection{Performance: Scale with Tensor Parallel}
\label{sec:expr:TP}
All previous experiments in this section examine the runtime improvement of \name{} in data parallel. 
In this experiment, we measure the runtime speedup of \name{} in the setting of combined data parallel (DP) and tensor parallel (TP).
We perform this experiment on Perlmutter and set TP to four, as each compute node has four A100s.
We use a GPT-2 model with 7B parameters, which is the maximum that can be hosted across four 40~GB A100s.
The scaling baseline is measured with one node.
Our test case runs on 32 nodes (i.e., 128 A100s).

From the measurements, we observe that \name{} runs 2.2x faster than AdamW on 128 A100s. 
The sustained scaling efficiency of \name{} is 73.4\% compared at 33.4\% of AdamW.
The results show that, in addition to data parallelism, \name{} is effective in reducing runtime cost with complex parallelism for LLM pretraining.

%We explore the scaling limit of the model size when Pier uses the mixture of DP and TP. We test it by setting the TP rank to 4, since Perlmutter has 4 A100s in a single compute node. We uses a self-modified model to mock a GPT-2 model with 7B parameters, which is the model size limit of our approach with TP size is 4 A100s (40GB). Figure~\ref{fig:tp} shows our result can achieve a speedup at 2.2x.

\begin{figure}[ht]
  \centering
  \includegraphics[width=1\columnwidth]{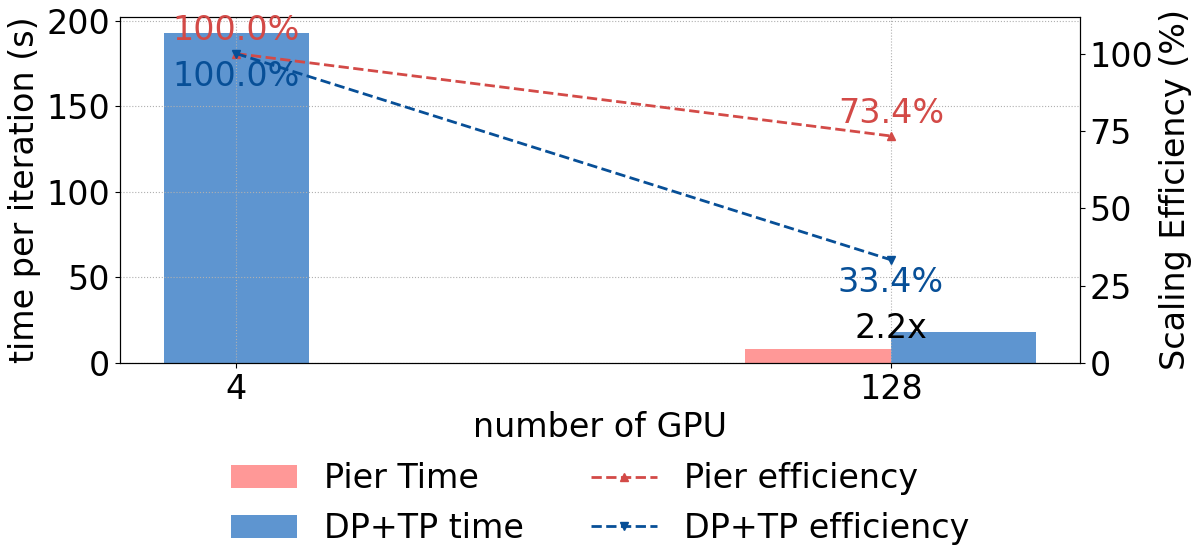}
  %\vspace{0.1in}
  \caption{\fontsize{9pt}{11pt}\selectfont
    Runtime and Scaling Efficiency Comparison between DP+TP and Pier. With a TP rank of 4, our approach achieves 2.2x speedup when scaling to 32x more GPUs.}
  \label{fig:tp}
  %\vspace{0.1in}
\end{figure}

\section{Conclusion}
We present \name{}, an optimization framework for LLM pretraining with relaxed global communication.
\name{} is enabled by the momentum warmup and momentum decay algorithms, and an efficient and scalable system design to address complex parallel strategies in LLM pretraining.
We verify the convergence properties of \name{} and identify the effectiveness boundary of global batch size, number of groups, and synchronization intervals.
We measure the speedup of \name{} compared to the widely adopted AdamW optimizer across machines, scales, and models.
The results show that \name{} scan speed up LLM-pretraining by 2.2x-3.7x for GPT-2 XL on 256 A100s on Perlmutter.
\name{} also exhibits a 2.2x speedup for GPT-2 7B on 256 A100s.
On Vista, \name{} reduces the end-to-end time cost of GPT-2 XL by 18.3-46.8\% on 128 GH200s.
\name{} is a strong optimization framework candidate in LLM pretraining, as it achieves significant speedups while preserving model performance.

%This paper aims to reduce the communication overhead during pretraining based on DiLoCo. We focus on addressing two key problems in DiLoCo: lack of convergence evaluation in i.i.d settings, and lack of end-to-end improvement. Our proposed Pier approach significantly fixed the first problem by leverage the fast convergence speed of AdamW approach and stabilize the switching point. Our experiment provides sufficient real-time improvement measurement for the lateral problem. This work could significantly make DiLoCo-like methods being used in practice since the convergence is guaranteed and performance gaining is clear. Our limitation is that the communication in outer loop is not considered. 

\bibliographystyle{IEEEtran}
\bibliography{references}

\end{document}